# RockGPT: Reconstructing three-dimensional digital rocks from single two-dimensional slice from the perspective of video generation

Qiang Zheng[1] and Dongxiao Zhang[2,*]

[1]Intelligent Energy Laboratory, Frontier Research Center, Peng Cheng Laboratory, Shenzhen 518000, P. R. China

[2]School of Environmental Science and Engineering, Southern University of Science and Technology, Shenzhen 518055, P. R. China

* Correspondence to: zhangdx@sustech.edu.cn

## Abstract

Random reconstruction of three-dimensional (3D) digital rocks from two-dimensional (2D) slices is crucial for elucidating the microstructure of rocks and its effects on pore-scale flow in terms of numerical modeling, since massive samples are usually required to handle intrinsic uncertainties. Despite remarkable advances achieved by traditional process-based methods, statistical approaches and recently famous deep learning-based models, few works have focused on producing several kinds of rocks with one trained model and allowing the reconstructed samples to satisfy certain given properties, such as porosity. To fill this gap, we propose a new framework, named RockGPT, which is composed of VQ-VAE and conditional GPT, to synthesize 3D samples based on a single 2D slice from the perspective of video generation. The VQ-VAE is utilized to compress high-dimensional input video, i.e., the sequence of continuous rock slices, to discrete latent codes and reconstruct them. In order to obtain diverse reconstructions, the discrete latent codes are modeled using conditional GPT in an autoregressive manner, while incorporating conditional information from a given slice, rock type, and porosity. We conduct two experiments on five kinds of rocks, and the results demonstrate that RockGPT can produce different kinds of rocks with the same model, and the reconstructed samples can successfully meet certain specified porosities. In a broader sense, through leveraging the proposed conditioning scheme, RockGPT constitutes an effective way to build a general model to produce multiple kinds of rocks simultaneously that also satisfy user-defined properties.

# 1. Introduction

Digital characterization of rocks provides great benefits for investigating pore-scale flow in oil reservoirs or groundwater aquifers via the modeling method. After obtaining three-dimensional (3D) digitalized porous structures, the research about how the microscopic structure affects macroscopic properties, such as permeability, can be conducted more efficiently than traditional experimental methods. Due to the spatial stochasticity of porous structures, a detailed understanding of pore-scale flow and accurate structure-property mappings usually requires a large number of digital rock samples, with the aim to quantify intrinsic uncertainties.

Over the last few decades, the development of 3D imaging techniques, such as X-ray computed tomography (e.g., Micro-CT and Nano-CT) [1-3], focused ion beam and scanning electron microscope (FIB-SEM) [4,5] and helium-ion-microscope (HIM) [6-8], for ultra-high-resolution imaging, have been widely adopted to physically scan rocks and build their digital twins. Even though physical scanning can provide digital rock samples with the highest fidelity, it cannot guarantee the requirements of quantity for uncertainty analysis, due to certain factors, such as limited access to imaging instruments, potentially expensive cost, and insufficient expertise in rock sample preparation and image post-processing. To overcome these limitations, diversified mathematical methods have been developed for random reconstruction of digital rocks. Among them, process-based methods, statistical approaches, and deep learning-based models are the most broadly used categories.

Process-based methods reconstruct digital rocks by means of simulating rock formation processes, including sedimentation, compaction, and diagenesis. The random packing of grains with a specified radius was commonly adopted to mimic the compaction and solidification process of sedimentary rocks, such as sandstone [9,10]. Owing to the relatively regular grains, the synthetic structures generally possess good connectivity and flow properties. It is difficult, however, to apply such methods to those rocks, such as carbonate, whose diagenetic process are not always known.

Statistical approaches work on an array of pixels in a regular grid, and leverage statistical modeling to represent binary void-grain structures that adhere to certain geostatistical constraints. Among these methods, two-point statistics (TPS) and multi-point statistics (MPS) are the most commonly used ones. Quiblier [11] first adopted Gaussian random field, a representative of the TPS method, to reconstruct 3D porous media based on 2D thin section images. Adler et al. [12] employed the auto-correlation function for the reconstruction of Fontainebleau sandstone from 2D images. Even though the crucial step, i.e., the matching of the two-point correlation function, in TPS methods is relatively easy to implement, it cannot sufficiently characterize the porous structure, especially for those structures with multi-phase and anisotropic characteristics.

In contrast, MPS can extract local multi-point features by scanning a training image with a certain template, which contributes to the incorporation of high-order information, and thus it achieves a better reconstruction performance than TPS methods [13]. The early MPS approach is CPU-intensive since it needs to rescan the whole training image to predict a new pixel. To address this constraint, Strebelle [14] proposed a single normal equation simulation (SNESIM) method, which employed a searching tree to scan the training image only once, and thus markedly improved efficiency. This strategy, however, necessitates a great memory requirement, especially when a large template is used to capture the morphological information. To overcome this limitation, cross-correlation-based simulation (CCSIM) was developed by Tahmasebi et al. [15] to realize MPS in a more efficient manner. The key idea behind CCSIM is that it treats 3D reconstruction as a stack of a list of successive 2D slices, which significantly alleviates the memory burdens. Even though CCSIM can reduce memory load via layer-by-layer reconstruction, its initial version possesses certain deficiencies, which stimulated the subsequent improvements, including MS-CCSIM for multi-scale features extraction [16], iterative CCSIM for avoiding discontinuities between neighboring blocks [17,18], and the three-step sampling method (TSS) for improving reconstruction quality along the stacking direction [19].

In recent years, with easier access to big data and advanced super-computers, deep learning (DL) techniques have experienced dramatic advancements, and their successes in image synthesis have inspired considerable applications in digital rock reconstruction. Compared with the previous methods, such as process-based and statistical methods, DL-based approaches can extract image features automatically from data, and thus avoid the necessity for prior expertise and complex hand-crafted feature design. Among DL models, generative adversarial networks (GANs) [20] and variational auto-encoder (VAE) [21] have achieved the most popularity for digital rock reconstruction since they can capture the data distribution and reproduce it very well. Mosser et al. [22] first adopted GANs to learn 3D images of several kinds of porous structures, and successfully reconstructed random samples with morphological and flow properties maintained. Shams et al. [23] combined GANs with the auto-encoder to reconstruct multi-scale porous media, thus enabling GANs to predict inter-grain pores while the auto-encoder provides the inner-grain ones.

Apart from reconstructing rock samples by taking random noise as input, the 2D slice should also be considered to serve as an input of reconstruction models, since, in most practical applications, 2D slices are obtained through physical scanning. Zhang et al. [24] proposed a hybrid model composed of GAN and VAE to produce 3D porous structures based on 2D images. The GAN-VAE hybrid model enables the encoder in VAE to characterize the statistical and morphological information of the 2D input image, and then generates a low-dimensional vector for the generator in GAN. Feng et al. [25] developed a BicycleGAN framework for mappings from 2D slices to 3D structures, and this method was successfully tested on two statistically isotropic materials and a non-stationary material with high efficiency. Although the above works have realized 2D-to-3D mappings, they still require 3D training data, which may pose challenges to areas where 3D data are challenging to acquire. As a consequence, Anderson et al. [26] first designed a new method, named RockFlow, to synthesize 3D structures purely based on 2D training data, whose core assumptions are that rock sample images are isotropic along all dimensions, and linear interpolation in latent space produces semantic interpolation in image space. There is another work [27] that adopted linear interpolation in latent space for 3D reconstruction based on 2D cross-section images, and it was realized by progressive growing GAN [28]. In addition, Kench and Cooper [29] developed a SliceGAN framework, which sufficiently utilized information contained in cross-sectional micrographs to statistically reconstruct 3D samples based on a single representative 2D image.

Even though 3D reconstructions based on 2D slices using deep learning methods have recently witnessed substantial progress, almost all of the current approaches only incorporate information from given 2D/3D images, while failing to explicitly consider other important properties, such as porosity, which may constrain its potential to produce highly representative samples. In addition, almost all of the existing models are trained on one kind of rock, which means that, for example, a model trained on sandstone data cannot be utilized for carbonate reconstructions. As for the reason, the current methods lack an effective conditioning scheme to incorporate additional information apart from image data.

In the present study, inspired by the layer-by-layer reconstruction strategy employed in CCSIM, we treat 3D reconstruction as an analogy of video synthesis, in which continuous frames can be viewed as slices of porous structures that are interconnected along any direction. In this setting, we propose a method, named RockGPT, which was originally developed as VideoGPT by Yan et al. [30] for video generation, to reconstruct 3D samples based on a single 2D slice. The RockGPT is composed of vector-quantized VAE (VQ-VAE) [31] and conditional GPT [32], in which VQ-VAE is first utilized to encode the input slices as discrete latent codes, and then conditional GPT is employed to model these discrete latent codes in an autoregressive manner while incorporating certain additional information, such as given 2D slices, rock type and porosity, through a proposed conditioning scheme.

The remainder of this paper proceeds as follows. Section 2 introduces the RockGPT architecture and its two components. The experimental data used in this work are briefly introduced in section 3, and the results are presented in section 4. Conclusions are provided in section 5.

## 2. Methodology

### 2.1 Vector quantized variational auto-encoder (VQ-VAE)

In terms of generative modeling, deep learning models can be broadly divided into two categories. One group is explicitly likelihood-based methods, such as PixelCNNs [33] and VAE; and the other group is likelihood-free methods, such as GANs [34]. Considering that the task in this work is sequential generation, i.e., generate subsequent slices based on the initial single slice, autoregressive models belonging to the likelihood-based category are comparably more appropriate. Regarding autoregressive modeling, it will be more efficient to construct models in a down-sampling latent space without spatial-temporal redundancies than that at the atomic level of all pixels across space and time. Therefore, it seems that building an autoregressive model over latent codes obtained by the VAE encoder is a good choice. However, the latent codes in VAE are continuous vectors, which are not as suitable as discrete vectors for adapting to cutting-edge autoregressive models, e.g., Transformer [35]. In addition, discrete representations are potentially a more natural fit than continuous features for many of the modalities in which we are interested. For instance, language is inherently discrete, and images are also composed of pixels whose values range from 0 to 255 with discrete integers. Therefore, it is of important meaning to learn discrete latent representations from high-dimensional input data, such as video, and then build an autoregressive model over the discrete latent codes in order to sample them and generate new realizations. VQ-VAE, which was developed by Oord et al. [31], can successfully accomplish this task.

As shown in Fig. 1, the VQ-VAE model takes a slice sequence $x$ as input, which will be passed through an encoder to produce output $z_e(x)$. The $z_e(x)$ is continuous vectors, and in order to obtain posterior categorical distribution $q(z|x)$ of discrete latent random variables $z$, we need to define a latent embedding space $e \in \mathcal{R}^{K \times D}$, where $K$ represents the size of the discrete latent space (i.e., a $K$-way categorical), and $D$ is the dimensionality of each embedding vector $e_i$. It can be seen that there are $K$ embedding vectors $e_i \in \mathcal{R}^D, i = 1, 2, \cdots, K$. Suppose that we have $n$ vectors with $D$ dimensionality in $z_e(x)$, and we should traverse the embedding space $e$ to find the nearest embedding vector $e_k$ for each component of $z_e(x)$. Then, each vector in $z_e(x)$ will find their nearest neighbors from the embedding space, and their index $k$ will compose the discrete latent variables $z$. Therefore, the categorical distribution $q(z|x)$ can be defined as one-hot as follows:

$$q(z = k|x) = \begin{cases} 1 & \text{for } k = \underset{j}{\operatorname{argmin}} \|z_e(x) - e_j\|_2 \\ 0 & \text{otherwise} \end{cases}. \tag{1}$$

Since discrete latent variable $z$ cannot feed the decoder directly, we replace it with the corresponding embedding vector by taking $z$ as an index of embedding space. Thus, the discrete embedding $z_q(x)$ is defined as below:

$$z_q(x) = e_k, \quad \text{where } k = \underset{j}{\operatorname{argmin}} \|z_e(x) - e_j\|_2, \tag{2}$$

where $z_q(x)$ has the same dimensionality as $z_e(x)$, but the former is more compressed than the latter, and consequently the subsequent sampling and reconstruction based on latent space will be more efficient. Finally, the discrete embedding $z_q(x)$ will serve as an input of the decoder to produce new or reconstructed videos. The network architecture of the encoder and decoder are illustrated in Appendix A.

Regarding the model training, the parameters coming from encoder ($\mathcal{E}$), decoder ($\mathcal{D}$) and embedding space ($e$) can be trained using the following objective:

$$\mathcal{L} = \|x - \mathcal{D}(z_q(x))\|_2^2 + \|sg[z_e(x)] - e\|_2^2 + \beta\|sg[e] - z_e(x)\|_2^2, \tag{3}$$

where $sg$ represents stop-gradient. The total loss $\mathcal{L}$ contains three terms, and the first term is reconstruction loss, which is utilized to train the encoder and decoder. Since Equation (2) cannot provide real gradients for back propagation, we copy gradients $\nabla_z \mathcal{L}$ from decoder input $z_q(x)$ to encoder output $z_e(x)$, as shown in the red line in Fig. 1. Consequently, the embedding space cannot get updated via the reconstruction loss due to the straight-through gradient estimation from $z_q(x)$ to $z_e(x)$. In order to learn the embedding space, vector quantization (VQ), which is one of the simplest dictionary learning algorithms, can be used to move the embedding vector $e_i$ to the encoder output $z_e(x)$. In practice, it can be realized by enforcing the second term of Equation (3), and due to the $sg$ operator, which is defined as identity in forward computing and has zero derivatives in backward propagation, only the embedding space needs to be updated during back propagation. Furthermore, to ensure that the encoder commits to an embedding, a commitment loss, which is defined as the third term in Equation (3), is added to purely train the encoder via adopting the $sg$ operator on the embedding space. The coefficient $\beta$ was chosen as 0.25 in all experiments of [31] and validated to facilitate a good performance, and thus we also follow this setting in our work.

After training the encoder, the decoder and the embedding space, the VQ-VAE has been endowed to down-sample a high-dimensional input to low-dimensional discrete latent codes and reconstruct the original samples. However, until now, it has had no ability to create new samples; in order to achieve this, we need to build an autoregressive model on $q(z|x)$ so that we can sample it to produce new latent codes and the corresponding output videos. In the original work of VQ-VAE, this model was built with PixelCNN or WaveNet. In the present work, however, it will be replaced by a new cutting-edge autoregressive model, named GPT, which will be introduced in the next section. For additional theoretical details about VQ-VAE, interested readers can refer to [31].

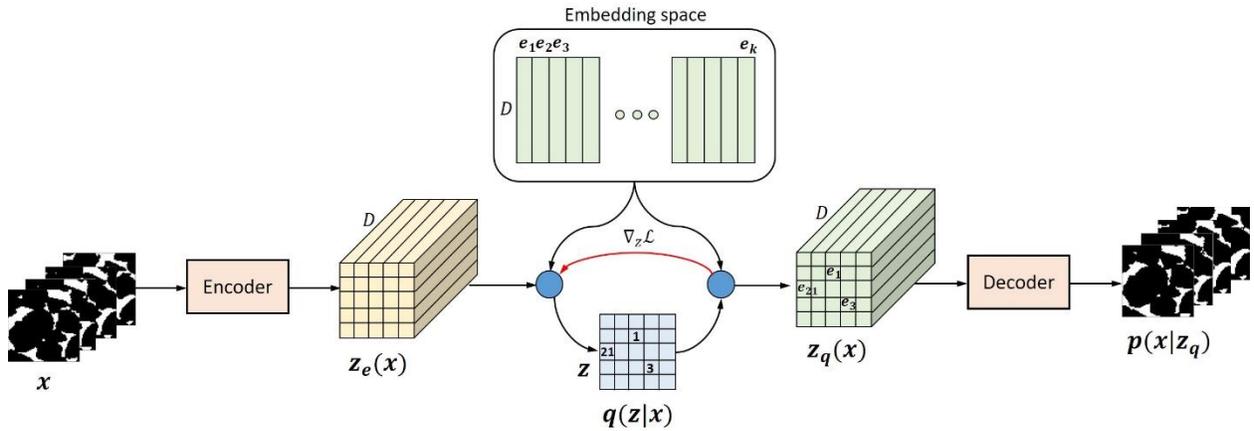

**Fig. 1.** The schematic of VQ-VAE. The encoder takes video-like data $x$, i.e., slice sequence, as an input, and outputs low-dimensional embedding vectors $z_e(x)$, which are continuous variables. In order to obtain discrete latent variables $z$, an embedding space $e$ needs to be learned via the vector quantization (VQ) algorithm, and then the discrete $z$ can be determined by a nearest neighbor look-up using embedding space. Afterwards, the discrete embedding $z_q(x)$ will be constructed with the embedding space by using discrete $z$ as an index of embedding vectors. Finally, the $z_q(x)$ will be taken as inputs of the decoder to produce reconstructed samples. The posterior categorical distribution $q(z|x)$ can be learned by an autoregressive model, such as PixelCNN, in order to sample it and generate new samples.

## 2.2 Conditional GPT

GPT is an abbreviation of Generative Pre-Training, which was created by Radford et al. [36] to build a language model in an unsupervised manner, followed by supervised fine-tuning on each specific task. GPT adopts Transformer as the model architecture [35], which has been demonstrated to achieve excellent performance in various fields, such as machine translation in the field of natural language processing [37,38], image classification and generation in the field of computer vision [32,39,40], and speech recognition and synthesis within audio-related applications [41,42]. This model choice enables GPT to provide a more structured memory and higher efficiency for handling long-term dependencies in sequence data due to the employment of the attention mechanism, compared to existing alternatives, such as recurrent neural networks.

Inspired by advances in unsupervised representation learning for natural language, Chen et al. [32] developed a similar model to learn useful representations for images, via training a GPT model to autoregressively predict image pixels without incorporating knowledge of image structures. Given the image data $m \in \mathcal{M}$ being flattened as a pixel sequence with length $n$, i.e., $m = (m_1, m_2, \cdots, m_n)$, the density $p(m)$ can be modeled autoregressively as follows:

$$p(m) = \prod_{i=1}^{n} p(m_i|m_1, m_2, \cdots m_{i-1}; \theta), \qquad (4)$$

where $\theta$ represents the network parameters. The model can be trained by minimizing the negative log-likelihood of the data:

$$\mathcal{L}_{\text{GPT}} = \mathop{\mathbb{E}}_{m \sim \mathcal{M}}[-\log p(m)]. \tag{5}$$

The Transformer decoder is adopted as the model architecture for GPT, which takes an input sequence $m_1, m_2, \cdots, m_n$ of discrete language tokens or image pixels, and produces a $d$-dimensional embedding for each position. As shown in Fig. 2, the GPT architecture is mainly composed of a stack of $N$ attention blocks. Each attention block has three sub-modules with residual connections, and each module takes layer normalization (LayerNorm) firstly [43] and dropout operation in the end. The difference lies in that the first two sub-modules adopt multi-head attention to compute dependencies within an input sequence, while the last sub-module utilizes fully-connected layers to gather information. As can be seen from Fig. 2, the discrete embeddings $z_q(x)$ produced by the trained VQ-VAE serve as the input of GPT, and the corresponding discrete latent codes $z$ play the role of category targets, which will be used to formulate the loss function (i.e., Equation (5)) on top of the output probabilities in each position. Therefore, we actually utilize Equation (4) to build an autoregressive model for discrete latent codes $z$. In order to encode the order of the sequence, the information about the relative or absolute position of elements in the sequence should be injected before the attention blocks by adding a positional encoding layer, which can be realized by enforcing a sine function on the position index [35].

Based on the original architecture of GPT, we add a condition block on it to build a conditional GPT, with an aim to incorporate some additional information, such as the given slice, rock type, and porosity. In particular, the conditioning of the initial slice is the key step that contributes to the 3D reconstruction based on one single slice. As shown in Fig. 2, when training the model, the first slice of the input sequence is chosen as the conditional slice, and its spatial features will be extracted by a ResNet block and a positional encoding layer. The architecture of ResNet used in this work is demonstrated in Appendix B. Regarding how to fuse the information extracted from the conditional slice, we perform cross-attention on the ResNet output using the second sub-module of each attention block, as illustrated by path ① in Fig. 2. In contrast, the conditional rock type or porosity are fairly low-dimensional data compared with the conditional slice, and we incorporate them by designing a conditional layer normalization for each sub-module of all attention blocks. It can be realized by enforcing a linear transformation, i.e., $y_{\text{new}} = y_{\text{original}} \times W + b$, whose gain ($W$) and bias ($b$) are parameterized as affine functions of the low-dimensional conditional vector, on the output ($y_{\text{original}}$) of the original layer normalization, as demonstrated by path ② in Fig. 2. Therefore, two types of conditioning, i.e., cross attention and conditional layer normalization, are utilized in this work to create a conditional GPT. For additional details about the attention mechanisms, i.e., self-attention and cross-attention, one can refer to [35].

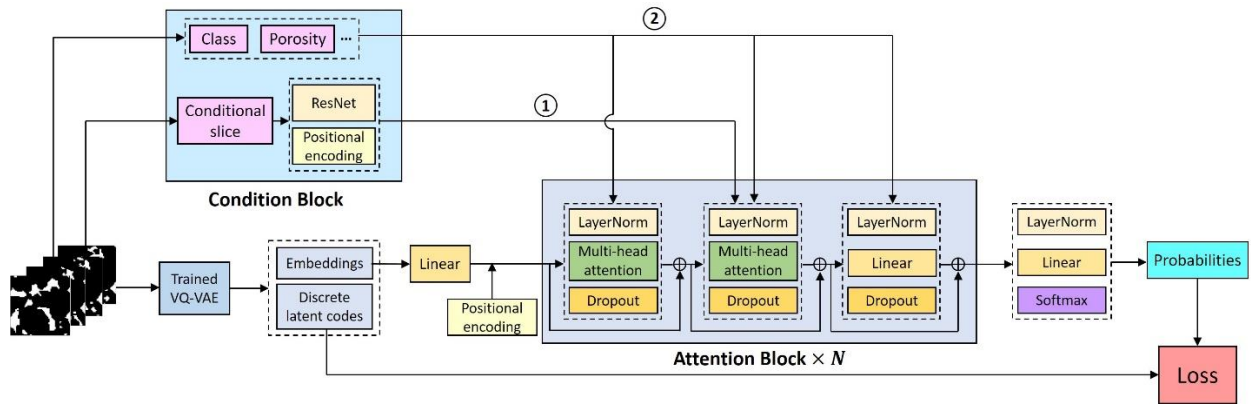

**Fig. 2.** The architecture of the conditional GPT. The GPT is mainly composed of $N$ attention blocks, and each block has three sub-modules with residual connections, which are marked as ⊕. The trained VQ-VAE should be utilized here to take slice sequence $x$ as input, and produce discrete embeddings $z_q(x)$ and latent codes $z$. The embeddings will go through the attention blocks after a linear transformation and positional encoding, while the discrete latent codes will serve as targets to build the loss function together with the output category probabilities of each position. In order to incorporate conditional information, cross attention and conditional layer normalization (LayerNorm) are employed, as illustrated by path ① and ②, respectively. The former is used to fuse information from the high-dimensional conditional slice, while the latter is employed to incorporate low-dimensional information, such as rock class and porosity.

## 2.3 RockGPT

Inspired by the creation of VideoGPT for generating videos based on the initial frame, we apply this framework to 3D digital rock reconstruction based on one single 2D slice. Considering how to produce highly representative samples, such as structures satisfying given porosities, we explore the conditioning scheme that can be integrated into model training. Followed by the settings of VideoGPT, when training the GPT model, we adopt cross attention to realize the conditioning of the 2D slice, and leverage conditional layer normalization to incorporate relatively low-dimensional properties, such as rock type and porosity, as mentioned previously. Furthermore, the conditioning of rock type allows the model to learn several kinds of rocks simultaneously, which avoids the situation in which one model is developed for one rock, and seamlessly conserves computational expense. We call the proposed framework RockGPT, and it comprises VQ-VAE and conditional GPT.

As shown in Fig. 3, the training of RockGPT can be realized by two stages in order. In the first stage, a VQ-VAE should be trained to learn a discretized latent space from high-dimensional data and reconstruct them via the loss function defined in Equation (3). In order to produce diversified samples, in the second stage, the discrete latent codes can be flattened (i.e., path ① in Fig. 3) and modeled using a conditional GPT in an autoregressive manner while incorporating certain additional information. After training the conditional GPT, we can sample it guided by some given conditional information to produce a new sequence of latent codes, which will be sent to the trained decoder of VQ-VAE to generate new samples, as illustrated by path ② in Fig. 3. It is worth noting that the last slice of generated samples can be utilized as a new conditional slice for the subsequent generation; therefore, we can stack them in order to reconstruct new samples, which consequently have a much larger size than that of training ones along the stacking direction.

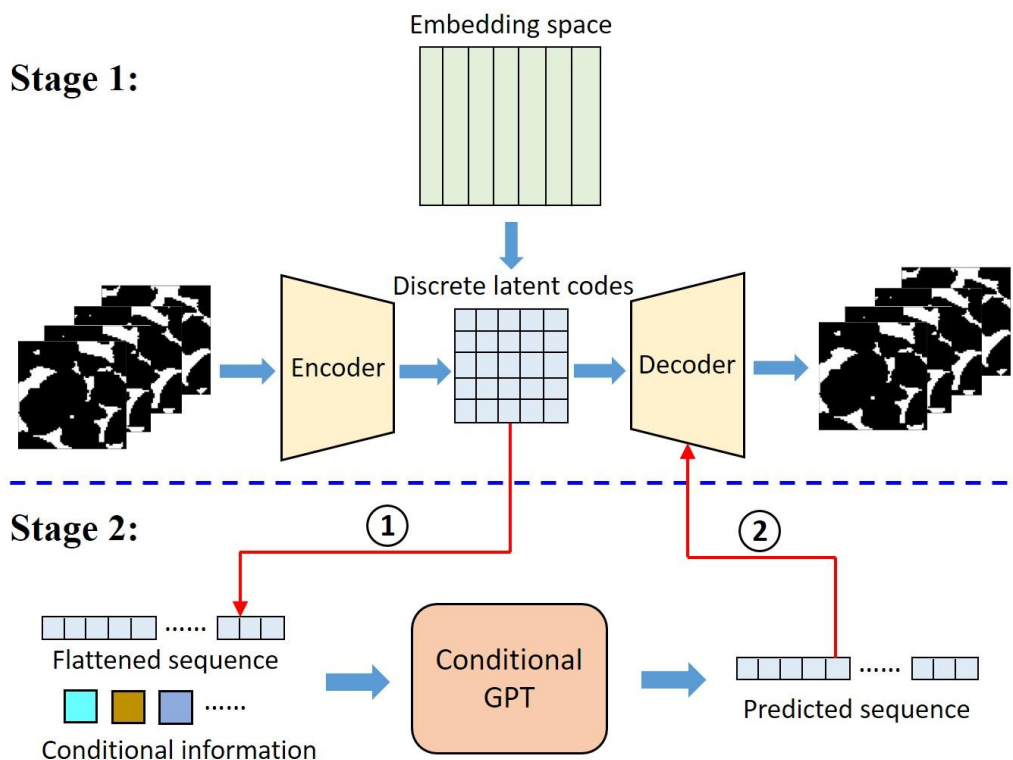

**Fig. 3.** The architecture of the proposed RockGPT. The training of RockGPT is divided into two stages. In the first stage, a VQ-VAE model is trained to compress high-dimensional slice sequences into a discretized latent space and reconstruct them. Then, the discrete latent codes will be flattened (i.e., path ①) and modeled with conditional GPT in the second stage, while incorporating some additional properties. For inference, a new latent sequence can be sampled from the trained GPT constrained by some given conditional information, and then the new latent sequence will be sent to the decoder in VQ-VAE for new slice generation (i.e., path ②).

## 3. Experimental data

In order to evaluate the reconstruction performance of the proposed RockGPT, we collect five kinds of rocks, i.e., Berea sandstone, Doddington sandstone, Estaillade carbonate, Ketton carbonate and Sandy multiscale medium, from public datasets, i.e., the Digital Rocks Portal (https://www.digitalrocksportal.org/), and their basic information is presented in Table 1. Considering that we have prepared a large number of samples of the five rock types with size $64^3$ voxels in our previous work [44], we can easily construct a new training dataset based upon it. In this work, the input of RockGPT is a sequence of slices, and it can be directly extracted with size $l \times 64 \times 64$ voxels from the original samples. Here, $l$ represents the length of the slice sequence, and it at least needs to satisfy the correlation length of porous structures, with an aim to capture the morphological characteristics of entire pores. However, it is not suggested to choose a large number for $l$ if available computing resources are limited, since memory usage will increase quadratically with the growth of $l$. In order to determine an appropriate $l$, we calculate two-point correlation functions for the original samples used in our previous work, and adopt an exponential model, i.e., $R(x) = \exp(-x/\lambda)$, to fit these functions for obtaining the correlation length $\lambda$ and calculating their averages. As shown in Table 1, Ketton carbonate has the largest mean correlation length with size 5.68 voxels, and consequently in this study, we set $l = 8$ voxels that can meet the requirements of correlation length of all rock types. Eventually, the slice sequences with size 8×64×64 voxels can be extracted from the original training dataset utilized in our previous work along the first axis with a spacing of four voxels, which means that an original sample with size $64^3$ voxels will reproduce five slice sequences with size 8×64×64 voxels. From the extracted dataset, 60,000 training samples and 500 testing samples are randomly selected for each rock type.

**Table 1.** Basic information of the five kinds of rock samples.

| Rock type | Original size (voxels) | Original resolution (μm) | Sample resolution (μm) | Mean correlation length (voxels) | Reference |
|---|---|---|---|---|---|
| Berea sandstone | $1000^3$ | 2.25 | 9.00 | 2.19 | [45] |
| Doddington sandstone | $700^3$ | 5.40 | 15.12 | 4.24 | [46] |
| Estaillade carbonate | $650^3$ | 3.31 | 8.60 | 4.17 | [47] |
| Ketton carbonate | $1000^3$ | 3.00 | 12.00 | 5.68 | [48] |
| Sandy multiscale medium | $512^3$ | 3.00 | 6.14 | 5.32 | [49] |

## 4. Results

### 4.1 Multi-rock reconstruction conditioned on single slice and rock type

In this case, we test the capability of RockGPT in producing 3D digital rocks of different kinds based on a single conditional slice. We select three kinds of rocks from Table 1, i.e., Berea sandstone, Doddington sandstone and Ketton carbonate, to conduct this experiment. As mentioned in the last section, the shape of the slice sequence is 8×64×64 voxels, and consequently the size of input for VQ-VAE should be formulated as $n \times 1 \times 8 \times 64 \times 64$ due to the utilization of 3D convolution. Here, $n$ represents the batch size and it is chosen as 32 in this work, and 1 is the number of channels. We set the shape of embedding space $e \in \mathcal{R}^{K \times D}$ in VQ-VAE as 1024×256, i.e., $K = 1024$ and $D = 256$. Meanwhile, the size of latent $z_e(x)$ is determined as $n \times 256 \times 2 \times 32 \times 32$ voxels, and it will affect the structural design of the encoder and decoder in VQ-VAE, which are listed in Appendix A. In addition, the latent size above means that for each sample, we have 2×32×32, i.e., 2,048 in total, vectors with dimensionality 256 to compute distance with the vectors in embedding space $e$, and thus the length of discrete $z$ is 2,048 voxels if it is flattened. In terms of model training, we firstly train the VQ-VAE for 50,000 iterations using the Adam optimizer [50] with the learning rate as $3e - 4$, which lasts for approximately 4.5 hours in four Tesla V100 GPUs. After obtaining the trained VQ-VAE, we continue to train the conditional GPT, for which the length of the input sequence, i.e., 2,048, has already been determined by the latent $z$ obtained from VQ-VAE. Concerning the conditional information, the initial slice with size $n \times 1 \times 1 \times 64 \times 64$ voxels of each input sequence is chosen as the conditional slice, and the conditional rock types are assigned as one-hot codes, i.e., [1, 0, 0] for Berea sandstone, [0, 1, 0] for Doddington sandstone, and [0, 0, 1] for Ketton carbonate. We train the conditional GPT for 100,000 iterations with the same optimizer as that in VQ-VAE, and it requires approximately 23 hours in four Tesla V100 GPUs.

After training the model, we can sample the conditional GPT autoregressively with given rock types being incorporated simultaneously to obtain a new sequence of discrete latent codes, and it will be translated into the corresponding discrete embeddings by looking up the embedding space in the trained VQ-VAE. Then, the trained decoder will take the discrete embeddings as input to generate new slice sequences. Since the length of the slice sequence is determined as eight voxels in this case, this means that we can only generate seven subsequent slices once based on the initial slice. Therefore, in order to generate samples with a larger size, the last slice can be set as a new conditional slice for the next sample production, and this operation can be iterated several times. For instance, if we want to generate samples with size $64^3$ voxels, we can repeat the above process nine times and concatenate the outputs of each time in order. It should be noted that the first slice of each generated sequence is also the preset conditional slice, and thus it should be omitted to avoid repetition when concatenating the outputs after the second iteration.

From the prepared testing dataset mentioned in section 3, we randomly select 100 samples for each rock type, and employ their first slices along the first axis as the conditional slice to generate new samples with size $64^3$ voxels. For metrics comparison, we also select real samples with the same amount and size as the fake ones from the original dataset, which is used for creating the training dataset of this work, as mentioned in section 3. As shown in Fig. 4, one real sample and three synthetic samples are randomly chosen for each rock type to demonstrate the reconstruction performance. It should be noted that the $x$-axis direction is used for stacking, and it can be seen that the pores along the stacking direction are highly connected, and the entire structures are visually and geologically realistic. To quantitatively evaluate the accuracy of conditioning on rock type, we adopt the metric Fréchet Video Distance (FVD) [51] to measure the distance between the samples of two rock types. Overall, the smaller is the FVD value, the more similar will be the two sample clusters. We utilize the real and synthetic samples prepared above to calculate FVD, and the results are presented in Table 2. It is obvious that only the combination of the fake data and real data of the same rock type can have a small FVD, which indicates that the conditioning of rock type has successfully guided the model to produce samples of specified rock type.

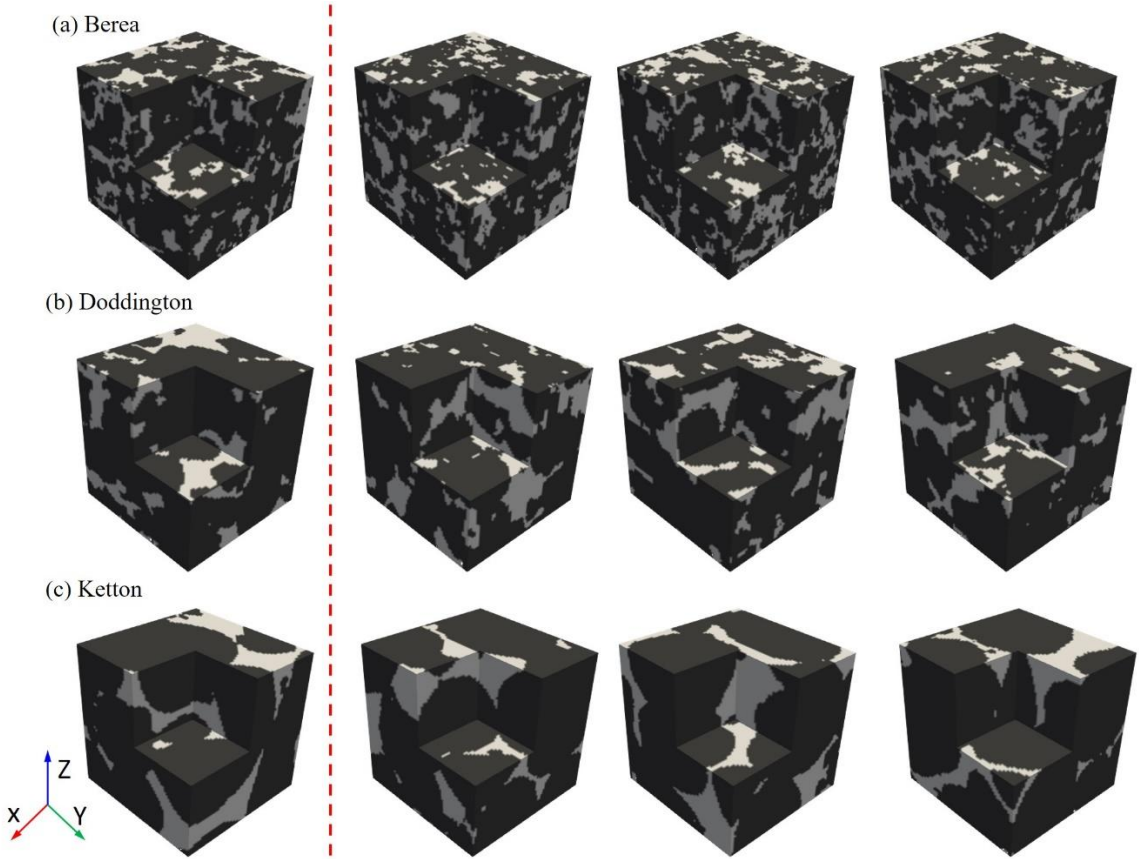

**Fig. 4.** The real samples (the column on the left of the red dashed line) and the synthetic samples (the columns on the right of the red dashed line) of (a) Berea sandstone, (b) Doddington sandstone, and (c) Ketton carbonate. All of the samples are with size $64^3$ voxels.

**Table 2.** Fréchet Video Distance (FVD) between the fake and real dataset with sample size 100 for each kind of rock.

| Real dataset  Fake dataset | Berea | Doddington | Ketton |
|---|---|---|---|
| Berea | **176.5** | 5712.8 | 5903.2 |
| Doddington | 4287.0 | **355.4** | 1231.9 |
| Ketton | 5575.8 | 1192.9 | **84.8** |

Even though visual realism and conditioning of rock type have been preliminarily validated, we adopt geostatistical analysis to further assess the reconstruction performance. The rock media used in this case can be expressed as a binary field as follows:

$$F(\mathbf{x}) = \begin{cases} 1 & \mathbf{x} \in \Omega_{\text{pore}} \\ 0 & \mathbf{x} \in \Omega_{\text{solid}} \end{cases}, \quad (6)$$

where $\mathbf{x}$ denotes any point of the rock image; and $\Omega_{\text{pore}}$ and $\Omega_{\text{solid}}$ represent pore and grain space, respectively. The first- and second-order moment of $F(\mathbf{x})$ can be used to characterize the structure of rocks, and they are defined as below:

$$\phi_F = \overline{F(\mathbf{x})}, \quad (7)$$

$$R_F(\mathbf{x}, \mathbf{x} + \mathbf{r}) = \frac{\overline{[\phi_F - F(\mathbf{x})] \cdot [\phi_F - F(\mathbf{x} + \mathbf{r})]}}{\phi_F - \phi_F^2}, \quad (8)$$

where $\phi_F$ is the porosity; and $R_F(\mathbf{x}, \mathbf{x} + \mathbf{r})$, termed the normalized two-point correlation function, is the probability that two points $\mathbf{x}$ and $\mathbf{x} + \mathbf{r}$, separated by lag distance $\mathbf{r}$, are located in the pore space $\Omega_{\text{pore}}$. In addition, we also employ Minkowski functionals, i.e., specific surface area and Euler characteristic, to evaluate morphological realism. The specific surface area is written as follows:

$$S_a = \frac{1}{V} \int dS, \quad (9)$$

where integration occurs at the solid-pore interface $S$; and $V$ is bulk volume. The Euler characteristic is defined as below:

$$\chi_V = \frac{1}{4\pi V} \int \frac{1}{r_1 r_2} dS, \quad (10)$$

where $r_1$ and $r_2$ are the principal radii of curvature of the solid-pore interface. To compute the above two Minkowski functionals, we utilize the BoneJ plugin [52] in the open-source software Fiji [53].

We also use the samples prepared above, i.e., 100 real and synthetic samples with size $64^3$ voxels for each rock type, to compute these metrics. As shown in Fig. 5(a), the porosity ranges of synthetic samples agree well with those of real samples for all kinds of rocks. The results of the normalized two-point correlation function in three directions are presented in Fig. 6. It can be seen that both the means and ranges of the two-point correlation function of generated samples can match well with those of real samples in any direction. It is worth noting that since the stacking along the $x$-axis direction in this work is based on sequences of slices, rather than the slice itself, and the length of the sequence can meet the requirements of correlation length, the discontinuity along the stacking direction, which occurs in traditional layer-by-layer reconstruction methods (e.g., CCSIM) [54], can be avoided in this work.

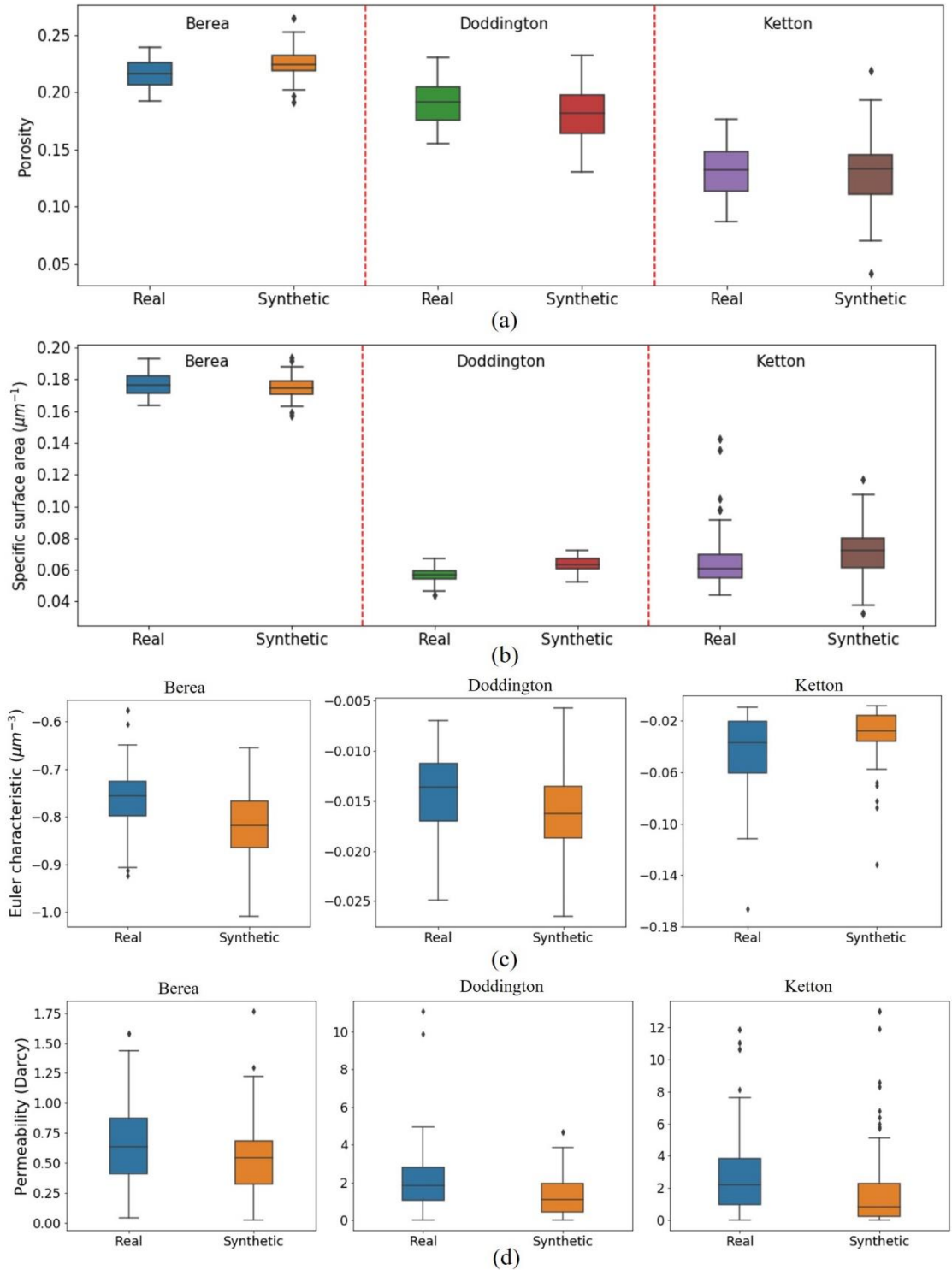

**Fig. 5.** Comparisons of (a) porosity, (b) specific surface area, (c) Euler characteristic and (d) permeability of real and synthetic samples for Berea sandstone, Doddington sandstone, and Ketton carbonate.

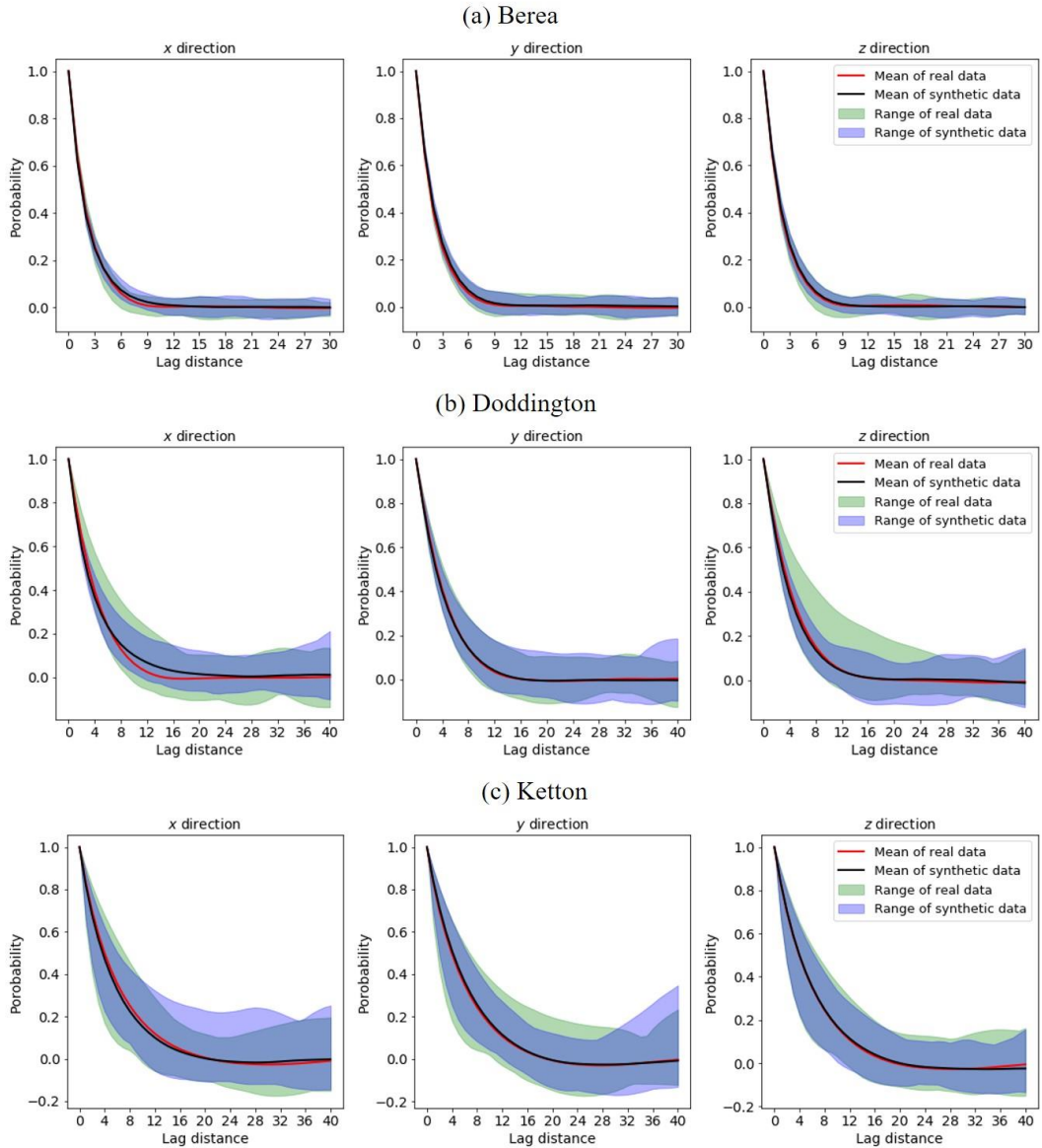

**Fig. 6.** Comparisons of the normalized two-point correlation of real and synthetic samples for (a) Berea sandstone, (b) Doddington sandstone, and (c) Ketton carbonate.

The matching performance of specific surface area and Euler characteristic are illustrated in Fig. 5(b) and Fig. 5(c), respectively, and it indicates that the synthetic samples can reproduce the real ones very well in terms of these two metrics. To evaluate the physical accuracy of the generated samples, we compute their absolute permeability using the single-phase Lattice Boltzmann method [55]. It can be seen from Fig. 5(d) that the permeability of fake samples has almost the same magnitude order and range as those of real samples for each rock type. Through the comparisons of two geostatistical metrics, two Minkowski functionals and one physical property, we can essentially conclude that the proposed RockGPT is able to reconstruct 3D samples based on a single slice of a specified rock type while maintaining its geological and physical realism.

### 4.2 Reconstruction conditioned on single slice, rock type, and porosity

After validating the conditioning performance of a given slice and rock type in the last section, we proceed to enable porosity as another conditional label in this case, with the aim to produce random samples that satisfy a specified porosity. Similar to rock type, porosity information is incorporated through the conditional layer normalization, and the conditioning is realized by enforcing the following two equations in order:

$$y_{\text{LN}} = y_{\text{LN}}^{\text{ini}} \times g_{\text{class}} + b_{\text{class}}, \qquad (11)$$

$$y_{\text{LN}} = y_{\text{LN}} \times g_{\text{prop}} + b_{\text{prop}}, \qquad (12)$$

where $y_{\text{LN}}^{\text{ini}}$ is the original output of layer normalization, which is in an unconditional setting; $g_{\text{class}}$ and $b_{\text{class}}$ are gain and bias, respectively, and they are obtained by applying a linear transformation on conditional rock type; and $g_{\text{prop}}$ and $b_{\text{prop}}$ are similar to above, but obtained from conditional porosity. Since porosity is a continuous value, which is different from the one-hot encoded rock type, we assign a new transformation (i.e., Equation (12)) to incorporate its information, rather than concatenating it with rock type. Therefore, the four coefficients, i.e., $g_{\text{class}}$, $b_{\text{class}}$, $g_{\text{prop}}$ and $b_{\text{prop}}$, play the role of carriers that take the conditional rock type and porosity into the GPT model. We select three kinds of rocks from Table 1, i.e., Doddington sandstone, Estaillade carbonate and Sandy multiscale medium, to test the conditioning performance, especially porosity, since the conditioning of rock type has been validated in the previous section. The training settings are the same as those for the last case, i.e., 50,000 iterations for training VQ-VAE and 100,000 for training conditional GPT, and consequently the training time is also similar to that of the last case.

With the trained model, we can generate new random 3D samples with specified porosity based on the given 2D slices of any rock type. In this case, we set different porosity targets for different rock types, i.e., $\phi_{\text{target}} = 0.21$ for Doddington sandstone, $\phi_{\text{target}} = 0.15$ for Estaillade carbonate, and $\phi_{\text{target}} = 0.34$ for Sandy multiscale medium. As in the previous section, we also generate 100 samples with size $64^3$ voxels for each rock type, and randomly select real samples of the same amount and size to make a comparison. As shown in Fig. 7, the porosities of synthetic samples can not only gather around the preset targets, but also have quite smaller ranges than those of the real dataset. These results indicate that the conditioning of porosity is successful.

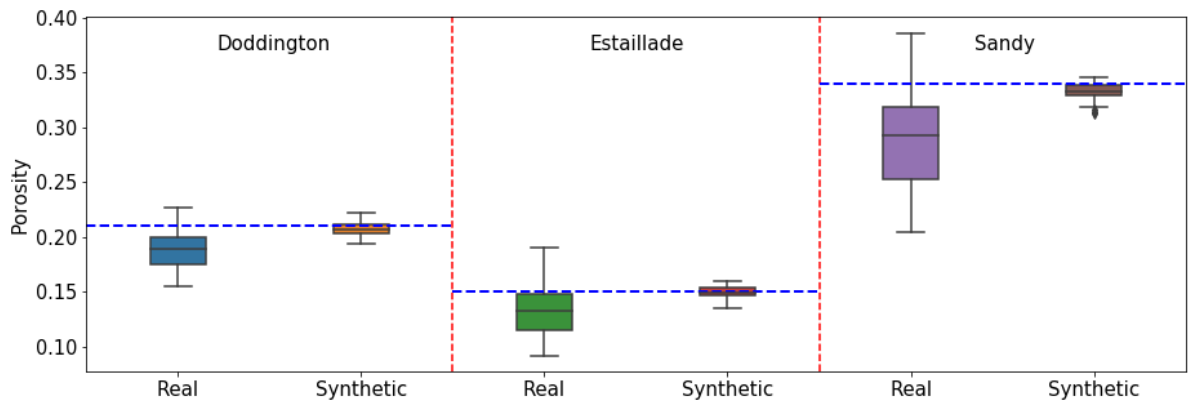

**Fig. 7.** Porosity of real and synthetic samples with specified porosity targets, i.e., $\phi_{\text{target}} = 0.21$ for Doddington sandstone, $\phi_{\text{target}} = 0.15$ for Estaillade carbonate, and $\phi_{\text{target}} = 0.34$ for Sandy multiscale medium.

In order to demonstrate the conditioning performance of porosity more vividly, we fix the conditional slice and modify the porosity target, with an aim to elucidate how the porous structure evolves along the stacking direction under different constraints of porosities. Here, we set two porosity targets for each rock type, i.e., $\phi_{\text{target}}^{\text{low}} = 0.16$ and $\phi_{\text{target}}^{\text{high}} = 0.22$ for Doddington sandstone, $\phi_{\text{target}}^{\text{low}} = 0.10$ and $\phi_{\text{target}}^{\text{high}} = 0.17$ for Estaillade carbonate, and $\phi_{\text{target}}^{\text{low}} = 0.21$ and $\phi_{\text{target}}^{\text{high}} = 0.36$ for Sandy multiscale medium. As shown in Fig. 8, the left column represents three fixed conditional slices for three rock types, whose porosities are denoted as $\phi_{\text{cond}}$, and the right eight columns are eight synthetic slices. Each rock type occupies two rows, and they stand for different porosity targets, with the upper rows having smaller targets, and the lower ones having larger targets. It can be clearly seen from the figure that the slices from left to right are highly interconnected, but exhibit a dissimilar trend of porosity. The pores of slices in the upper rows have a distinct shrinkage trend, while those in the lower rows show a slight expansion, since all of the $\phi_{\text{cond}}$ have a larger distance to $\phi_{\text{target}}^{\text{low}}$ than to $\phi_{\text{target}}^{\text{high}}$. Moreover, we continue to produce samples with size $64^3$ voxels based on the results of Fig. 8, and we calculate their porosities and permeabilities. As shown in Table 3, the porosities of synthetic samples are markedly close to the targets with small relative errors, and the samples with larger porosity targets also have larger permeabilities, which is as expected. Therefore, we can validate that the porosity has been successfully incorporated into the model, and can effectively affect the morphology of the porous structure while maintaining its visual and geological realism.

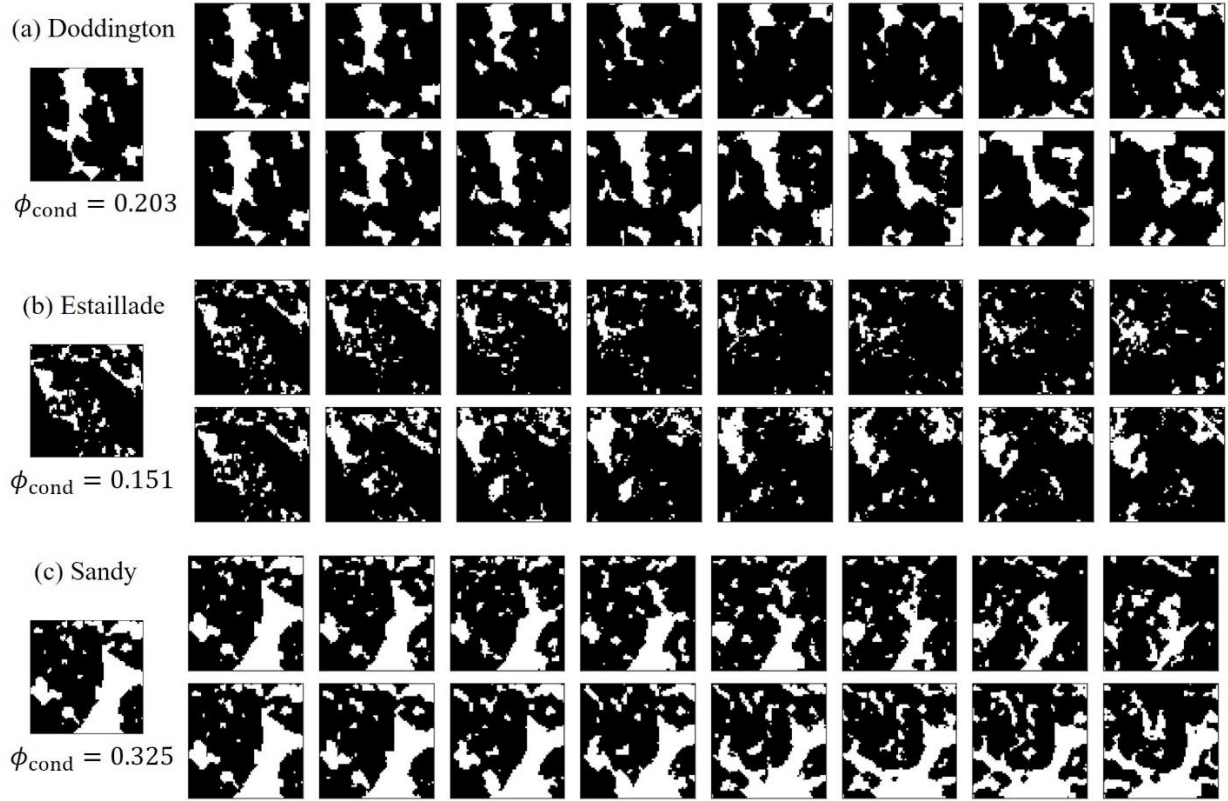

**Fig. 8.** Generated samples (eight continuous slices) with two kinds of porosity targets based on single slices (the left column), whose porosities are marked as $\phi_{cond}$. The preset porosity targets for (a) Doddington are 0.16 (the 1st row) and 0.22 (the 2nd row), (b) Estaillade are 0.10 (the 3rd row) and 0.17 (the 4th row), and (c) Sandy are 0.21 (the 5th row) and 0.36 (the 6th row).

**Table 3.** Porosity and permeability of generated samples with size $64^3$ voxels using the same targets as those in Fig. 8.

| Rock type | Porosity (relative error) | Permeability (Darcy) |
|---|---|---|
| Doddington | 0.156 (2.50%) | 1.16 |
|  | 0.212 (3.63%) | 3.42 |
| Estaillade | 0.094 (3.00%) | 0.36 |
|  | 0.166 (2.35%) | 1.30 |
| Sandy | 0.207 (1.43%) | 3.76 |

|                |        |
|----------------|--------|
| 0.355 (1.38%)  | 11.45  |

## 5. Conclusion

Inspired by the layer-by-layer reconstruction utilized in geostatistical methods, e.g., CCSIM, and its similarity with video synthesis, which is usually realized in a frame-by-frame manner, we propose to reconstruct 3D rock samples from 2D slices in the way of video generation by leveraging cutting-edge deep learning models. Therefore, in this work, we proposed a framework, named RockGPT, which is composed of VQ-VAE and conditional GPT, for 3D digital rock reconstruction based on a single 2D cross-section image. In the RockGPT architecture, VQ-VAE should be trained first to provide a down-sampled latent space, which was discretized and thus beneficial for autoregressive modeling, and a reconstruction path from discretized embeddings to the original video space. After obtaining the trained VQ-VAE, the conditional GPT could be trained to model the discrete latent codes in an autoregressive manner, in which cross-attention and conditional layer normalization were adopted to realize the conditioning. It should be noted that the conditioning of the slice, which was realized by the cross-attention mechanism, constituted the key point for 2D-to-3D mappings in this study, and it also enabled us to produce larger samples than the training ones, since the generated samples could also provide new conditional slices to make the generation proceed continuously. In addition, the conditional layer normalization contributed to the conditioning of rock type, which made it possible to produce multiple kinds of rocks with a single model instead of one model being developed for only one kind of rock, and effortlessly conserved computational resources. The conditioning of porosity was realized in the same manner as rock type, and it enabled us to produce 3D samples that satisfy the given porosity, which will provide certain benefits for downstream researches, such as fluid flow modeling in porous media with controllable porosities.

Even though RockGPT realized 3D reconstruction based on 2D slices through a novel perspective, i.e., video synthesis, it also possessed certain limitations that need to be addressed in the future. First, the GPT model was built on the attention mechanism, which is sensitive to the length of the input sequence and will induce a quadratic memory increase with sequence length, and consequently the latent size cannot be very large. At the same time, in order to preserve sufficient information in latent space and guarantee the reconstruction performance, the size of discrete latent codes in VQ-VAE cannot be very small. Therefore, a trade-off exists with respect to latent size between the VQ-VAE and GPT, which was sufficiently discussed in [30]. The sparse Transformer [56], which is able to handle relatively longer sequences with local and strided attention, can be a possible method to overcome this issue. Another limitation is that, even though we can produce samples with a larger size than that of training ones in the stacking direction, we still cannot enlarge the size along the other two directions in the current framework. To generate larger samples, one can consider combining RockGPT and conditional GANs utilized in our previous work [44]. Specifically, one can train RockGPT first to produce 3D samples based on a given slice, and then employ conditional GANs to learn the generated samples. Finally, the 3D samples with a larger size can be reconstructed by the trained conditional GANs.

## Acknowledgments

This work is partially funded by the Shenzhen Key Laboratory of Natural Gas Hydrates (Grant No. ZDSYS20200421111201738), the SUSTech - Qingdao New Energy Technology Research Institute, and the China Postdoctoral Science Foundation (Grant No. 2020M682830).

# Appendix A: The network architecture of the encoder and decoder in VQ-VAE

In this work, the sequence of slices with size 8×64×64 voxels is taken as input of the encoder, which is mainly composed of a down-sampling block and several attentional residual blocks. The down-sampling ratio determines how many 3D convolutions are needed for the down-sampling block. For instance, if we want to obtain a latent space with size 2×32×32 voxels from the input sequence, we need two down-sampling 3D convolutions with stride as two voxels, whose layer settings can be seen in Table A1. In the end of the down-sampling block, we add a 3D convolution that does not modify the shape. The attention residual block is designed as shown in Fig. A1, where we adopt batch normalization (BN) and axial attention layers following [57]. The 3D convolutions used in the attention residual block only modify the number of output channels, while keeping the size of feature maps unchanged. In this study, there are four attention residual blocks being utilized, followed by a BN and ReLU operation to final outputs.

The structure of the decoder is almost symmetric to that of the encoder, and the biggest difference is that the down-sampling block in the encoder is replaced with the up-sampling block in the decoder. As shown in Fig. A1(b), the discretized embedding firstly goes through a 3D convolution to change the number of channels. Subsequently, four attention residual blocks are adopted, which share the same settings as those in the encoder. Finally, the up-sampling block is used to map the feature space into the original video space. The number of transposed 3D convolution (ConvT3D) is also determined by the up-sampling rate, with each ConvT3D enlarging the side length as two times that of the previous ones. The settings of kernel size, stride, and corresponding output shape are listed in Table A1 and A2 for the encoder and decoder, respectively.

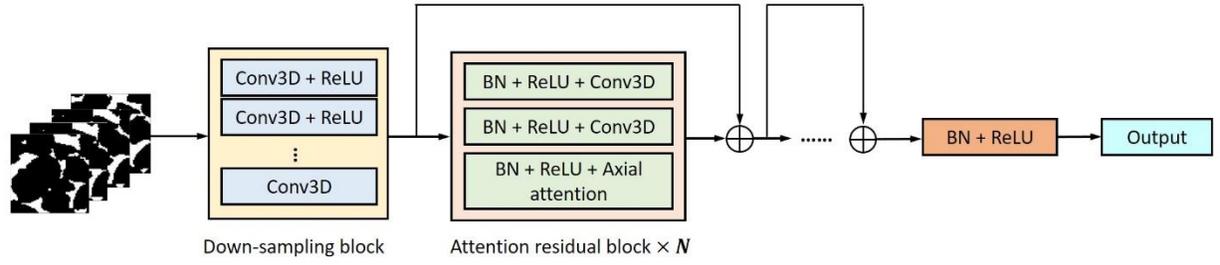

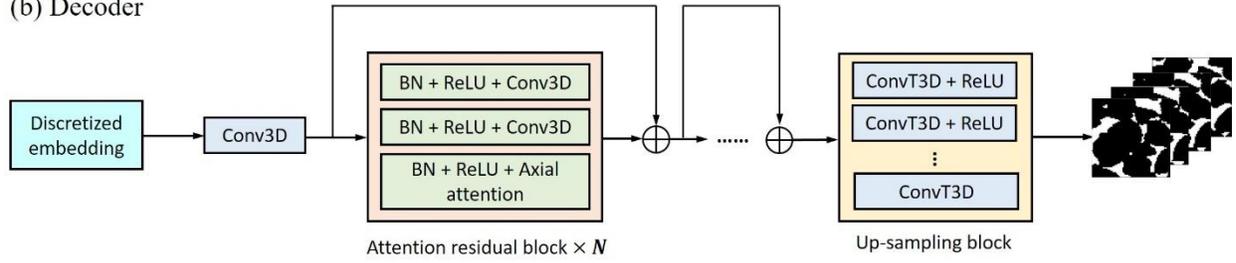

**Fig. A1.** Network architecture of (a) the encoder and (b) the decoder of VQ-VAE.

**Table A1.** Layer settings of the encoder in VQ-VAE.

| Block | Layer | Kernel size | Stride | Output shape | Trainable parameters |
|---|---|---|---|---|---|
| Down-sampling block | Conv3D | 4×4×4 | 2×2×2 | $n$ ×240×4×32×32 | 15,600 |
| | ReLU | - | - | - | - |
| | Conv3D | 4×4×4 | 2×1×1 | $n$ ×240×2×32×32 | 3,686,640 |
| | ReLU | - | - | - | - |
| | Conv3D | 3×3×3 | 1×1×1 | $n$ ×240×2×32×32 | 1,555,440 |
| Attention residual block 1 | BN+ReLU | - | - | - | 480 |
| | Conv3D | 3×3×3 | 1×1×1 | $n$ ×128×2×32×32 | 829,440 |
| | BN+ReLU | - | - | - | 256 |
| | Conv3D | 1×1×1 | 1×1×1 | $n$ ×240×2×32×32 | 30,720 |
| | BN+ReLU | - | - | - | 480 |
| | Axial attention | - | - | - | 691,920 |
| Attention residual block 2 | BN+ReLU | - | - | - | 480 |
| | Conv3D | 3×3×3 | 1×1×1 | $n$ ×128×2×32×32 | 829,440 |
| | BN+ReLU | - | - | - | 256 |
| | Conv3D | 1×1×1 | 1×1×1 | $n$ ×240×2×32×32 | 30,720 |
| | BN+ReLU | - | - | - | 480 |
| | Axial attention | - | - | - | 691,920 |
| Attention residual block 3 | BN+ReLU | - | - | - | 480 |
| | Conv3D | 3×3×3 | 1×1×1 | $n$ ×128×2×32×32 | 829,440 |
| | BN+ReLU | - | - | - | 256 |
| | Conv3D | 1×1×1 | 1×1×1 | $n$ ×240×2×32×32 | 30,720 |
| | BN+ReLU | - | - | - | 480 |
| | Axial attention | - | - | - | 691,920 |

| Block | Layer | Kernel size | Stride | Output shape | Trainable parameters |
|---|---|---|---|---|---|
| Attention residual block 4 | BN+ReLU | - | - | - | 480 |
| | Conv3D | 3×3×3 | 1×1×1 | $n \times 128 \times 2 \times 32 \times 32$ | 829,440 |
| | BN+ReLU | - | - | - | 256 |
| | Conv3D | 1×1×1 | 1×1×1 | $n \times 240 \times 2 \times 32 \times 32$ | 30,720 |
| | BN+ReLU | - | - | - | 480 |
| | Axial attention | - | - | - | 691,920 |
| | BN+ReLU | - | - | - | 480 |
| Total trainable parameters | | | | | 11.5M |

**Table A2.** Layer settings of the decoder in VQ-VAE.

| Block | Layer | Kernel size | Stride | Output shape | Trainable parameters |
|---|---|---|---|---|---|
| | Conv3D | 3×3×3 | 1×1×1 | $n \times 240 \times 2 \times 32 \times 32$ | 1,659,120 |
| Attention residual block 1 | BN+ReLU | - | - | - | 480 |
| | Conv3D | 3×3×3 | 1×1×1 | $n \times 128 \times 2 \times 32 \times 32$ | 829,440 |
| | BN+ReLU | - | - | - | 256 |
| | Conv3D | 1×1×1 | 1×1×1 | $n \times 240 \times 2 \times 32 \times 32$ | 30,720 |
| | BN+ReLU | - | - | - | 480 |
| | Axial attention | - | - | - | 691,920 |
| Attention residual block 2 | BN+ReLU | - | - | - | 480 |
| | Conv3D | 3×3×3 | 1×1×1 | $n \times 128 \times 2 \times 32 \times 32$ | 829,440 |
| | BN+ReLU | - | - | - | 256 |
| | Conv3D | 1×1×1 | 1×1×1 | $n \times 240 \times 2 \times 32 \times 32$ | 30,720 |
| | BN+ReLU | - | - | - | 480 |
| | Axial attention | - | - | - | 691,920 |
| Attention residual block 3 | BN+ReLU | - | - | - | 480 |
| | Conv3D | 3×3×3 | 1×1×1 | $n \times 128 \times 2 \times 32 \times 32$ | 829,440 |
| | BN+ReLU | - | - | - | 256 |
| | Conv3D | 1×1×1 | 1×1×1 | $n \times 240 \times 2 \times 32 \times 32$ | 30,720 |
| | BN+ReLU | - | - | - | 480 |
| | Axial attention | - | - | - | 691,920 |
| Attention residual block 4 | BN+ReLU | - | - | - | 480 |
| | Conv3D | 3×3×3 | 1×1×1 | $n \times 128 \times 2 \times 32 \times 32$ | 829,440 |
| | BN+ReLU | - | - | - | 256 |
| | Conv3D | 1×1×1 | 1×1×1 | $n \times 240 \times 2 \times 32 \times 32$ | 30,720 |
| | BN+ReLU | - | - | - | 480 |
| | Axial attention | - | - | - | 691,920 |
| Up-sampling block | Conv3D | 4×4×4 | 2×2×2 | $n \times 240 \times 4 \times 64 \times 64$ | 3,686,640 |
| | ReLU | - | - | - | - |
| | Conv3D | 4×4×4 | 2×1×1 | $n \times 1 \times 8 \times 64 \times 64$ | 15,361 |
| Total trainable parameters | | | | | 11.6M |

# Appendix B: The network structure of ResNet for extracting information of the conditional slice

To extract the features from the conditional slice, we adopt a convolutional neural network with ResNet architecture in this work. As shown in Fig. B1, the conditional slice with size 1×64×64 voxels serves as an input for ResNet, and then goes through an initial block composed of 3D convolution, layer normalization (LN), and ReLU activation. Subsequently, four groups containing residual connections are employed for further representation learning, and each group has two blocks, which differ in the first two layers and the variables for residual connection, as illustrated in Fig. B1. The network composition for the four groups are identical, except for the kernel size and stride, which leads to a gradual down-sampling from group 1 to group 4. The kernel size, stride, and the number of trainable parameters are listed in Table B1. It should be noted that the first dimension of the slice needs to be padded since its length cannot be less than the kernel size. As a result, we pad zeros along the outside of the input slice to make its shape as 3×65×65 voxels when using 3D convolutions with kernel size as three voxels. Moreover, the LN and ReLU operation cannot modify the shape of feature maps, and thus we ignore it in Table B1.

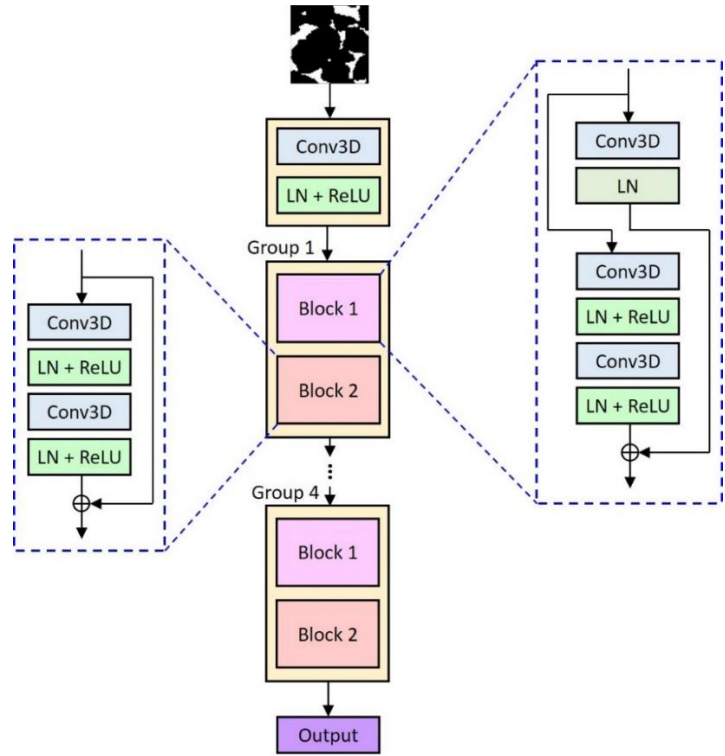

**Fig. B1.** Graphical representation of the network architecture of ResNet.

**Table B1.** Basic information of the ResNet architecture.

| Group | Block | Layer | Kernel size | Stride | Output shape | Trainable parameters |
|---|---|---|---|---|---|---|
| | Initial block | Conv3D | 3×3×3 | 1×1×1 | $n \times 64 \times 1 \times 64 \times 64$ | 1,728 |
| | | LN+ReLU | - | - | - | 128 |
| Group 1 | Block 1 | Conv3D | 1×1×1 | 1×1×1 | $n \times 64 \times 1 \times 64 \times 64$ | 4,096 |
| | | LN | - | - | - | 128 |
| | | Conv3D | 3×3×3 | 1×1×1 | $n \times 64 \times 1 \times 64 \times 64$ | 110,592 |
| | | LN+ReLU | - | - | - | 128 |
| | | Conv3D | 3×3×3 | 1×1×1 | $n \times 64 \times 1 \times 64 \times 64$ | 110,592 |
| | | LN+ReLU | - | - | - | 128 |
| | Block 2 | Conv3D | 3×3×3 | 1×1×1 | $n \times 64 \times 1 \times 64 \times 64$ | 110,592 |
| | | LN+ReLU | - | - | - | 128 |
| | | Conv3D | 3×3×3 | 1×1×1 | $n \times 64 \times 1 \times 64 \times 64$ | 110,592 |
| | | LN+ReLU | - | - | - | 128 |
| Group 2 | Block 1 | Conv3D | 1×1×1 | 1×2×2 | $n \times 128 \times 1 \times 32 \times 32$ | 8,192 |
| | | LN | - | - | - | 256 |
| | | Conv3D | 3×3×3 | 1×2×2 | $n \times 128 \times 1 \times 32 \times 32$ | 221,184 |
| | | LN+ReLU | - | - | - | 256 |
| | | Conv3D | 3×3×3 | 1×1×1 | $n \times 128 \times 1 \times 32 \times 32$ | 442,368 |
| | | LN+ReLU | - | - | - | 256 |
| | Block 2 | Conv3D | 3×3×3 | 1×1×1 | $n \times 128 \times 1 \times 32 \times 32$ | 442,368 |
| | | LN+ReLU | - | - | - | 256 |
| | | Conv3D | 3×3×3 | 1×1×1 | $n \times 128 \times 1 \times 32 \times 32$ | 442,368 |
| | | LN+ReLU | - | - | - | 256 |
| Group 3 | Block 1 | Conv3D | 1×1×1 | 1×2×2 | $n \times 256 \times 1 \times 16 \times 16$ | 32,768 |
| | | LN | - | - | - | 512 |
| | | Conv3D | 3×3×3 | 1×2×2 | $n \times 256 \times 1 \times 16 \times 16$ | 884,736 |
| | | LN+ReLU | - | - | - | 512 |
| | | Conv3D | 3×3×3 | 1×1×1 | $n \times 256 \times 1 \times 16 \times 16$ | 1769,472 |
| | | LN+ReLU | - | - | - | 512 |
| | Block 2 | Conv3D | 3×3×3 | 1×1×1 | $n \times 256 \times 1 \times 16 \times 16$ | 1769,472 |
| | | LN+ReLU | - | - | - | 512 |
| | | Conv3D | 3×3×3 | 1×1×1 | $n \times 256 \times 1 \times 16 \times 16$ | 1769,472 |
| | | LN+ReLU | - | - | - | 512 |
| Group 4 | Block 1 | Conv3D | 1×1×1 | 1×1×1 | $n \times 576 \times 1 \times 16 \times 16$ | 147,456 |
| | | LN | - | - | - | 1,152 |
| | | Conv3D | 3×3×3 | 1×1×1 | $n \times 576 \times 1 \times 16 \times 16$ | 3,981,312 |
| | | LN+ReLU | - | - | - | 1,152 |
| | | Conv3D | 3×3×3 | 1×1×1 | $n \times 576 \times 1 \times 16 \times 16$ | 3,981,312 |
| | | LN+ReLU | - | - | - | 1,152 |
| | Block 2 | Conv3D | 3×3×3 | 1×1×1 | $n \times 576 \times 1 \times 16 \times 16$ | 3,981,312 |
| | | LN+ReLU | - | - | - | 1,152 |
| | | Conv3D | 3×3×3 | 1×1×1 | $n \times 576 \times 1 \times 16 \times 16$ | 3,981,312 |
| | | LN+ReLU | - | - | - | 1,152 |
| Total trainable parameters | | | | | | 24.3M |